\documentclass[11pt]{article}
\usepackage{amssymb}
\usepackage{amsfonts}
\usepackage{graphicx}
\usepackage{amsmath}
\usepackage{hyperref}

\makeatletter \@addtoreset{equation}{section} \makeatother

\newtheorem{theorem}{Theorem}[section]
\newtheorem{lemma}{Lemma}[section]

\newtheorem{remark}{Remark}[section]
\newtheorem{definition}{Definition}[section]
\newtheorem{proposition}{Proposition}[section]

\newcommand{\mdet}{\mathrm{det}}

\newcommand{\Tr}{\mathrm{Tr}\,}

\begin{document}
\title{Characteristic polynomials for  random band matrices near the threshold}
\author{
 Tatyana Shcherbina
\thanks{Department of Mathematics, Princeton University, Princeton, USA, e-mail: tshcherbyna@princeton.edu. Supported in part by NSF grant DMS-1700009.}}

\date{}
\maketitle

\begin{abstract}
The paper continues \cite{TSh:14}, \cite{SS:ChP} which study the behaviour of second correlation function of characteristic polynomials of the special case of
$n\times n$ one-dimensional Gaussian Hermitian random band matrices,  when the covariance of the elements is determined by the matrix
$J=(-W^2\triangle+1)^{-1}$. Applying the transfer matrix approach, we study the case when the bandwidth $W$ is proportional to the threshold $\sqrt{n}$.

\end{abstract}
\section{Introduction}
As in \cite{TSh:14}, \cite{SS:ChP}, we consider Hermitian $n\times n$ matrices $H$
  whose entries $H_{ij}$ are random
complex Gaussian variables with mean zero such that
\begin{equation}\label{ban}
\mathbf{E}\big\{ H_{ij}H_{lk}\big\}=\delta_{ik}\delta_{jl}J_{ij},
\end{equation}
where
\begin{equation}\label{J}
J_{ij}=\left(-W^2\Delta+1\right)^{-1}_{ij},
\end{equation}
and $\Delta$ is the discrete Laplacian on $\mathcal{L}=[1,n]\cap \mathbb{Z}$ with Neumann boundary conditions.
It is easy to see that  the variance of matrix elements $J_{ij}$ is exponentially small when $|i-j|\gg W$, and so  $W$ can be considered as the width of the band.

The density of states $\rho$ of the ensemble is given by the well-known Wigner semicircle law (see
\cite{BMP:91, MPK:92}):
\begin{equation}\label{rho_sc}
\rho(E)=(2\pi)^{-1}\sqrt{4-E^2},\quad E\in[-2,2].
\end{equation}
 Random band matrices (RBM) provide
a natural and important model to study eigenvalue statistic and quantum transport
in disordered systems as they interpolate between classical Wigner matrices, i.e. Hermitian
random matrices with all independent identically distributed elements, and random
Schr$\ddot{\hbox{o}}$dinger operators, where only a random on-site potential is present in addition to the
deterministic Laplacian on a regular box in $d$-dimension lattice.
Such matrices have various application in physics: the
eigenvalue statistics of RBM is in relevance in quantum chaos, the quantum dynamics
associated with RBM can be used to model conductance in thick wires, etc.

One of the main long standing problem in the field is to prove a  fundamental  physical conjecture formulated in late 80th (see \cite{Ca-Co:90}, \cite{FM:91}). 
The conjecture states that  the eigenvectors of $n\times n$ RBM are completely delocalized and the local  spectral  statistics  
governed  by  random  matrix (Wigner-Dyson) statistics  
for large bandwidth $W$, and by Poisson statistics for a small $W$ (with exponentially localized eigenvectors). The transition is conjectured to be sharp and for RBM 
in  one  spatial  dimension  occurs around  the  critical  value $W=\sqrt{n}$. This  is  the  analogue  of  the  celebrated  Anderson  
metal-insulator  transition  for random Schr$\ddot{\hbox{o}}$dinger operators.

The conjecture on the crossover in RBM with $W\sim\sqrt n$ is supported by physical derivation due to Fyodorov and Mirlin (see \cite{FM:91}) based on supersymmetric formalism, and also by the so-called Thouless scaling. However, there are only partial results on the mathematical level of rigour (see reviews \cite{Bour:rev}, 
\cite{SS:rev} and references therein for the details).

The only result that rigorously demonstrate the threshold around $W\sim\sqrt n$ for a certain eigenvalue statistics was obtain in
\cite{TSh:14} (regime $W\gg \sqrt{n}$), \cite{SS:ChP} (regime $W\ll \sqrt{n}$). 
Instead of eigenvalue correlation functions these papers deal with more simple object which is  the
second correlation functions of characteristic polynomials:
\begin{equation}\label{F_2k}
F_{2}(x_1,x_2)=\mathbf{E}\Big\{\mdet(x_1-H)\mdet(x_2-H)\Big\}.
\end{equation}
The main results of  \cite{TSh:14}, \cite{SS:ChP} concern the asymptotic behaviour of this function for
\begin{equation*}
x_{1,2}=E+\dfrac{\xi_{1,2}}{n\rho(E)},\quad E\in (-2,2), \quad \xi_1,\xi_2\in [-C,C].
\end{equation*}
Namely, let 
\begin{equation*}
D_2=F_2(E,E), \quad \bar F_2(x_1,x_2)=D^{-1}_2\cdot F_2(x_1,x_2).
\end{equation*}
Then we have the following theorem
\begin{theorem}[\textbf{\cite{TSh:14}, \cite{SS:ChP}}]\label{thm:old}
For the 1d RBM of (\ref{ban}) -- (\ref{J}) we have 
\begin{equation*}
\lim\limits_{n\to\infty} 
\bar F_{2}\Big(E+\dfrac{\xi}{2n\rho(E)},E-\dfrac{\xi}{2n\rho(E)}\Big)=
\left\{
\begin{array}{cc}
\dfrac{\sin \pi\xi}{\pi\xi},& W\ge n^{1/2+\theta};\\
1,&  1 \ll W\le \sqrt {\dfrac{n} {C_*\log n}},
\end{array}
\right.
\end{equation*}
where the limit is uniform in $\xi$ varying in any compact set $C\subset\mathbb{R}$. Here $E\in (-2,2)$, 
and $\rho(x)$ is defined in (\ref{rho_sc}).
\end{theorem}
The purpose of the present paper is to complete Theorem \ref{thm:old} by the study of correlation functions of characteristic polynomials (\ref{F_2k})
near the threshold $W\sim \sqrt{n}$. The main result is
\begin{theorem}\label{thm:main}
For the 1d RBM of (\ref{ban}) -- (\ref{J}) with $n=C_*W^2$ we have 
\begin{equation*}
\lim\limits_{n\to\infty} 
\bar F_{2}\Big(E+\dfrac{\xi}{2n\rho(E)},E-\dfrac{\xi}{2n\rho(E)}\Big)=
(e^{-C^*\Delta_U- i\xi\hat\nu}\cdot 1,1),
\end{equation*}
where $C^*=C_*/(2\pi\rho(E))^2$. In this formula  $(\cdot,\cdot)$ is an inner product on a 2-dimensional sphere $\mathbb{S}^2$, 
$\Delta_U$ is a Laplace  operator  on $\mathbb{S}^2$
\begin{align*}
\Delta_U=-\frac{d}{dx}x(1-x)\frac{d}{dx},\quad  x=|U_{12}|^2,
\end{align*}
$U$ is  a $2\times 2$ unitary matrix, and $\hat\nu$ is an operator of multiplication on
\begin{align}\label{nu}
 \nu(U)=1-2|U_{12}|^2
\end{align}
on $\mathbb{S}^2$. 
\end{theorem}
\begin{remark} It is easy to see that if $W\gg \sqrt{n}$ (and so $C^*\to 0$), then we have
\[
(e^{-C^*\Delta_U- \pi i\xi\hat\nu}\cdot 1,1)\sim (e^{- \pi i\xi\hat\nu}\cdot 1,1)=\dfrac{\sin \pi\xi}{\pi \xi}.
\]
Similarly if $W\ll \sqrt{n}$ (and so $C^*\to \infty$), then we get
\[
(e^{-C^*\Delta_U- \pi i\xi\hat\nu}\cdot 1,1)\sim (e^{-C^*\Delta_U}\cdot 1,1)=1.
\]
Thus the result of Theorem \ref{thm:main} "glue" together two parts of Theorem \ref{thm:old}.
\end{remark}
\begin{remark} The study of eigenfunctions and spectral statistics in the critical regime (near the threshold) is of independent interest.
Critical wave-functions at the point of the Anderson localization transition are
expected to be multifractal. Moreover, multifractal structure occurs in a critical regime
of power-law banded random matrices (see the review \cite{EM:08} and reference therein
 for the details). Although  the correlation functions of characteristic polynomials (\ref{F_2k}) are not
 reach enough to feel this phenomena, the techniques developed in the paper can be useful 
 in studying the usual correlation functions of 1d RBM near the threshold.
\end{remark}

The proof of Theorem \ref{thm:main} is based on the techniques of \cite{SS:ChP}.
Namely, we apply the version of transfer matrix approach introduced in \cite{SS:ChP} to the integral representation obtained in \cite{TSh:14} 
by the supersymmetry techniques (note that the integral representation does not contain Grassmann integrals, see Proposition \ref{p:repr}). 

The paper is organized as follows. In Section $2$ we rewrite $F_2$ as an action of the $n$-th degree of some transfer operator $K_\xi$ (see (\ref{K_xi}) below) 
and outline the proof of Theorem \ref{thm:main}. In Section $3$ we collect all preliminaries results obtained in \cite{SS:ChP}.
Section $4$ deals  with the proof of Theorem \ref{thm:main}. 

We denote by $C$, $C_1$, etc. various $W$ and $n$-independent quantities below, which
can be different in different formulas. To reduce the number of notations, we also use the same letters for the integral operators and their kernels.

\section{Outline of the proof of Theorem \ref{thm:main}}

First, we rewrite $F_2$ as an action of the $n-1$-th degree of some transfer operator, as it was done in \cite{SS:ChP}.
 
For $X\in \hbox{Herm}(2)$ define
\begin{align}\label{F_cal}
f:=\mathcal{F} (X)&=\exp\Big\{-
\frac{1}{4}\, \Tr\Big(X+\frac{i\Lambda_0}{2}\Big)^2+\frac{1}{2}\,\Tr \log
\big(X-i\Lambda_0/2\big)-C_+\Big\},\\ \notag
f_\xi:=\mathcal{F}_\xi (X)&=\mathcal{F} (X)\cdot \exp\Big\{-
\frac{i}{2n\rho(E)}\, \Tr X\hat\xi\Big\}
\end{align}
with $\hat\xi=\hbox{diag}\,\{\xi,-\xi\}$,  $\Lambda_0=E\cdot I_2$, 
\begin{align}\label{a_pm}
&a_\pm=\pm\sqrt{1-E^2/4}\\
\label{C_+}
&C_+=\frac{1}{4}\, \Tr\Big(a_+I+\frac{i\Lambda_0}{2}\Big)^2-\frac{1}{2}\,\Tr \log
\big(a_+I-i\Lambda_0/2\big).
\end{align}
Set also $\mathcal{H}=L_2[\hbox{Herm}(2)]$, and let $K, K_\xi: \mathcal{H}\to\mathcal{H}$ be operators with the kernels 
\begin{align} \label{K}
K(X,Y)&=\dfrac{W^4}{2\pi^2}\,\mathcal{F}(X)\,\exp\Big\{-\frac{W^2}{2}\Tr
(X-Y)^2\Big\}\,\mathcal{F}(Y);\\
\label{K_xi}
K_{\xi}(X,Y)&=\dfrac{W^4}{2\pi^2}\,\mathcal{F}_\xi(X)\,\exp\Big\{-\frac{W^2}{2}\Tr
(X-Y)^2\Big\}\,\mathcal{F}_\xi(Y).
\end{align}
As it was proved in \cite{SS:ChP}, Section 2, we have
\begin{proposition}[\textbf{\cite{SS:ChP}}]\label{p:repr}
The second correlation function of characteristic polynomials of (\ref{F_2k}) for 1D Hermitian Gaussian band
matrices (\ref{ban}) -- (\ref{J}) can be represented as follows:
\begin{align}\label{F_rep}
F_2\Big(E+\dfrac{\xi}{n\rho(E)},E-\dfrac{\xi}{n\rho(E)}\Big)=-C_n(\xi)\cdot W^{-4n}\mdet^{-2} J\cdot ( K^{n-1}_\xi f_{\xi},\bar f_{\xi}),
\end{align}
where $(\cdot,\cdot)$ is a standard inner product in $\mathcal{H}$, $\rho$ is defined in (\ref{rho_sc}), and
\begin{equation*}
C_n(\xi)=\exp\big\{2nC_++\xi^2/n\rho(E)^2\big\}
\end{equation*}
with $C_+$ of (\ref{C_+}).
\end{proposition}

For arbitrary compact operator $M$  denote by $\lambda_j(M)$ the $j$th (by its modulo) eigenvalue
of $M$, so that $|\lambda_0(M)|\ge|\lambda_1(M)|\ge\dots$.

The idea of the transfer operator approach is very simple and natural.
 Let $\mathcal{K}(X,Y)$ be the matrix kernel of the compact  integral operator in $\oplus_{i=1}^pL_2[X,d\mu(X)]$. Then 
\begin{align*}\notag
&\int g(X_1) \mathcal{K}(X_1,X_2)\dots \mathcal{K}(X_{n-1},X_n)f(X_n)   \prod d\mu(X_i)=(\mathcal{K}^{n-1}f,\bar g)\\
&=\sum_{j=0}^\infty\lambda_j^{n-1}(\mathcal{K})c_j,\quad with\quad
c_j=(f,\psi_j)(g,\tilde\psi_j),
\end{align*}
where $\psi_j$ are  eigenvectors corresponding to $\lambda_j(\mathcal{K})$, and $\tilde \psi_j$ are the eigenvectors of $\mathcal{K}^*$. Hence, to study the correlation function,
it suffices to study the eigenvalues and eigenfunctions of the integral operator  with the kernel $\mathcal{K}(X,Y)$.

The main difficulties  in application of this approach to  (\ref{F_rep}) are
the complicated structure and non self-adjointness of the corresponding transfer operator $K_\xi$ of (\ref{K_xi}). 

In fact, since the analysis of eigenvectors of  non self-adjoint operators is rather involved, it is  simpler  to work with the resolvent 
analog of (\ref{F_rep})
\begin{align}\label{res_rep}
(K_\xi^{n-1}f_\xi,\bar f_\xi)=-\frac{1}{2\pi i}\oint_{\mathcal{L}}z^{n-1}(G_\xi(z)f_{\xi},\bar f_{\xi})dz,\quad 
G_\xi(z)=(K_\xi-z)^{-1},
\end{align}
where $\mathcal{L}$ is any closed contour which enclosed all eigenvalues of $K_\xi$. 

To explain the idea of the proof, we start from the definition
\begin{definition}\label{def:1}
 We shall say that the operator $\mathcal{A}_{n,W}$  is equivalent to  $\mathcal{B}_{n,W}$  ($\mathcal{A}_{n,W}\sim\mathcal{B}_{n,W}$) on some contour
 $\mathcal{L}$ if
\[ \int_{\mathcal{L}}z^{n-1}((\mathcal{A}_{n,W}-z)^{-1}f,\bar g)dz= \int_{\mathcal{L}}z^{n-1}((\mathcal{B}_{n,W}-z)^{-1}f,\bar g) dz \, (1+o(1)),\quad n,W\to \infty,\]
with some $f,g$ depending of the problem.
\end{definition}
The idea  is to find  some operator equivalent to $ K_\xi$  whose spectral analysis we are ready to perform.

It is easy to see that the stationary  points of the function $\mathcal{F}$ of  (\ref{F_cal})  are
\begin{align}\label{st_points_1}
X_+&=a_+\cdot I_2,\quad X_-=a_-\cdot I_2;\\
X_\pm(U)&=a_+\,ULU^*,\quad U\in \mathring{U}(2), \notag
\end{align}
where $a_\pm$ is defined in (\ref{a_pm}), $ \mathring{U}(2):=U(2)/U(1)\times U(1)$, $L=\hbox{diag}\,\{1,-1\}$.
Notice also that the value of $|\mathcal{F}|$ at points (\ref{st_points_1}) is $1$.

Roughly speaking, the first step in the proof of Theorem \ref{thm:main}  is to show that if we introduce the projection $P_{s}$  onto  the $W^{-1/2}\log W$-neighbourhoods of the saddle points $X_+$, $X_-$ and the saddle "surface" $X_\pm$, then in the sense of Definition \ref{def:1}
\begin{align*}
K_{\xi}\sim P_{s}K_{\xi}P_{s}=:K_{m,\xi}.
\end{align*}
To study the operator $K_{m,\xi}$  near the saddle ``surface" $X_\pm$ we  use the  "polar coordinates". Namely,
introduce
\begin{align}\label{t}
t=(a_1-b_1)(a_2-b_2), \quad p(a,b)=\dfrac{\pi}{2}(a-b)^2,
\end{align}  
and denote by $dU$ the integration with respect to the Haar measure
on the group $\mathring{U}(2)$: in the standard parametrization 
\begin{equation}\label{U_param}
U=\left(
\begin{array}{cc}
\cos \varphi&\sin\varphi\cdot e^{i\theta}\\
-\sin\varphi\cdot e^{-i\theta}&\cos\varphi
\end{array}
\right),
\end{equation}
we have
\[
dU=\dfrac{1}{\pi}u\,du\, d\theta, \quad u=|\sin\varphi|\in [0,1], \quad \theta \in [0,2\pi).
\]
Consider the space $L_2[\mathbb{R}^2,p]\times L_2[\mathring{U}(2),dU]$.
The inner product and the action of an integral operator in this space are
\begin{align*}
&(f,g)_p=\int f(a,b)\bar g(a,b)p(a,b)\,da\,db;\\
&(Mf)(a_1,b_1,U_1)=\int M(a_1,b_1,U_1;a_2,b_2,U_2)\,f(a_2,b_2,U_2)\,p(a_2,b_2)da_2\,db_2\,dU_2.
\notag\end{align*} 
Changing the variables
\[X=U^*\Lambda U,\quad \Lambda=\mathrm{diag}\{a,b\},\quad a>b,\quad U\in \mathring{U}(2),\]
we obtain that $K_{\xi}=K+\widetilde{K}_\xi$ can be  represented as an integral operator in $L_2[\mathbb{R}^2,p]\times L_2[\mathring{U}(2),dU]$  defined by the kernel
\begin{align}\label{rep_2}
&K_\xi(X,Y)=K(a_1,a_2,b_1,b_2,U_1,U_2)+\widetilde K_\xi(a_1,a_2,b_1,b_2,U_1,U_2),
\end{align}
where
\begin{align} \notag
&K(a_1,a_2,b_1,b_2,U_1,U_2)=t^{-1}A(a_1,a_2)A(b_1,b_2)K_*(t,U_1,U_2);\\ \label{K(U)}
&K_*(t,U_1,U_2):=W^2t\cdot e^{tW^2\Tr U_1U_2^*L(U_1U_2^*)^*L/4-tW^2/2};\\
&\widetilde K_\xi(a_1,a_2,b_1,b_2,U_1,U_2)=K(a_1,a_2,b_1,b_2,U_1,U_2)\big(e^{(\nu(a_1-b_1,U_1)+\nu(a_2-b_2,U_2))/n}-1\big);\notag\\
&\nu(x,U)=-\frac{i\xi\,x}{4\rho(E)}\,\Tr ULU^*L.
\notag\end{align}
$K_*$ here is a contribution of the unitary group $\mathring{U}(2)$, and $\nu(x,U)$ is a perturbation of $\mathcal{F}$ appearing in $\mathcal{F}_\xi$ (see (\ref{F_cal})).
Operator $A$ is a contribution of eigenvalues $a,b$ that has the form
\begin{align}
\label{A}
&A(x,y)=(2\pi)^{-1/2}We^{-g(x)/2}e^{-W^2(x-y)^2/2}e^{-g(y)/2}, \\ \notag
&g(x)=(x+iE/2)^2/2-\log (x-iE/2)-C_+;
\end{align}
Note also that
\begin{equation}\label{b_Ktil}
\|\widetilde K_\xi\|\le C/n
\end{equation}
with some absolute $C>0$

The main properties of $K_*$ are given in the following proposition:
  \begin{proposition}\label{p:K(U)}
If we  consider
 $K_*(t,U_1,U_2)$ of (\ref{K(U)})  as a kernel of the self-adjoint integral operator in $L_2[\mathring{U}(2), dU]$, then 
  its eigenvectors  $\{\phi_{\bar j}(U)\}$ ($\bar j=(j,s)$, $j=0,1,\ldots$, $s=-j,\ldots, j$) do not depends on $t$ and are the standard spherical harmonics: 
\begin{equation*}
\phi_{j,s}(U)=l_{j,s}\, P_j^s(\cos 2\varphi)\, e^{is\theta}=l_{j,s} \Big(\dfrac{d}{dx}\Big)^s P_j(x)\Big|_{x=1-2|U_{12}|^2}(2\bar{U}_{11} U_{12})^s,
\end{equation*}
where $U$ has the form (\ref{U_param}), and $P_j^s$ is an  associated Legendre polynomial
\begin{align*}
&P_j^s(\cos x)=(\sin x)^s  \Big(\dfrac{d}{d\cos x}\Big)^s P_j(\cos x),\quad P_j(x)=\dfrac{1}{2^jj!}\dfrac{d^j}{dx^j}(x^2-1)^j,\\ \notag
&l_{j,s}=\sqrt{\dfrac{(2j+1) (j-s)!}{(j+s)!}}.
\end{align*}
Moreover, the subspace $L_2[u, dU]\subset L_2[\mathring{U}(2), dU]$ of the functions depending on $\varphi$ only is invariant under
$K_*$, and the restriction of $K_*$ to $L_2[u, dU]$ 
has eigenvectors
\begin{equation}\label{phi_j0}
\phi_{j}(U):=\phi_{j,0}(U).
\end{equation}
  The corresponding
 eigenvalues $\{\lambda_{j}(t)\}_{j=0}^\infty$, if  $t>d>0$, where $d$ is some absolute positive constant, have the form
 \begin{align}\label{l_j}
&\lambda_{0}(t)=1-e^{-W^2t},\\ \notag
&\lambda_{j}(t)=(1-e^{-W^2t})\Big(1-\frac{j(j+1)}{W^2t}(1+O(j^2/W^2t)\Big).
\end{align}
\end{proposition}
Notice that since $$ \Tr U^*LUL=2(1-2u^2),$$
functions $\mathcal F$, $\mathcal{F}_\xi$ do not depend on $\theta$ of (\ref{U_param}), and hence according to Proposition \ref{p:K(U)}
in what follows we can consider the restriction of $K$, $K_*$ and $\widetilde K_\xi$ of (\ref{K(U)}) to $L_2[u, dU]$ (to simplify notations we will denote these
restriction by the same letters).

In addition, it follows from Proposition \ref{p:K(U)} that if we introduce the following basis in $L_2[\mathbb{R}^2,p]\times L_2[u,dU]$
\begin{align*}
&\Psi_{\bar k,j}(a,b,U)=\Psi_{\bar k}(a,b)\phi_{j}(U), \\
&\Psi_{\bar k}(a,b)=\sqrt{\dfrac{2}{\pi}}\,(a-b)^{-1}\psi_{k_1}(a)\psi_{k_2}(b),
\notag\end{align*}
where $\bar k=(k_1,k_2)$, and $\{\psi_{k}(x)\}_{k=0}^\infty$ is a certain basis in $L_2[\mathbb{R}]$, then the matrix of
$K$ of (\ref{K(U)}) in this basis has a ``block diagonal  structure", which means that
\begin{align}\label{block}
&(K\Psi_{\bar k',j},\Psi_{\bar k, j_1})_p=0,\quad j\not=j_1\\
&(K\Psi_{\bar k',j},\Psi_{\bar k, j})_p=(K_j\Psi_{\bar k'},\Psi_{\bar k})_p\notag\\
=&\int \lambda_{j}(t)A(a_1,a_2)A(b_1,b_2)\psi_{k_1}(a_1)\psi_{k_2}(b_1)
\psi_{k_1'}(a_2)\psi_{k_2'}(b_2)da_1db_1da_2db_2.
\notag\end{align}
The next step in the proof of Theorem {\ref{thm:main}} is to show that only the neighbourhood of the saddle "surface" $X_\pm$ gives the main contribution to the integral, and moreover we can restrict the number of $\phi_j$ to $l=[\log W]$. More precisely, we are going to show that
in the sense of Definition \ref{def:1}
\begin{equation}\label{step2}
K_{m,\xi}\sim \mathcal{P}_{l}K_{m,\xi}\mathcal{P}_{l}=:K_{m,l,\xi},
\end{equation}
where $\mathcal{P}_l$ is the projection on the linear span of $\{\Psi_{\bar k,j}(a,b,U)\}_{j\le l, |\bar k|\le m}$. 

For the further resolvent analysis we want to put  $t$ in the definition of $K_*$ and $a_1-b_1$, $a_2-b_2$ in the definition of $\widetilde K_\xi$
(see (\ref{t}), (\ref{rep_2}) -- (\ref{K(U)})) equal to their saddle-point value $t^*=(a_+-a_-)^2=4\pi^2\rho(E)^2$ and $a_+-a_-=2\pi\rho(E)$ correspondingly.
More precisely we want to show that 
in the sense of Definition \ref{def:1}
\begin{equation}\label{tens_pr}
K_{m,l,\xi}\sim \mathcal{A}_m\otimes K_{*\xi,l}
\end{equation}
where
\begin{align}\label{K_*xi}
&K_{*\xi,l}=Q_l\,K_{*\xi}\,Q_l,\\
\notag &K_{*\xi}(U_1,U_2)=W^2t^*\cdot e^{t^*W^2\Tr U_1U_2^*L(U_1U_2^*)^*L/4-t^*W^2/2}\cdot e^{n^{-1}\nu(2\pi\rho(E),U_1)+n^{-1}\nu(2\pi\rho(E),U_2)}
\end{align}
and $Q_l$ is the projection on  $\{\phi_j(U)\}_{j\le l}$. The operator $\mathcal{A}_m$ in (\ref{tens_pr}) is defined as
\begin{align}\label{A_m}
\mathcal{A}_m=P_mA(a_1,a_2) A(b_1,b_2) P_m,
\end{align}
where $P_m$ is the projection on $\{\Psi_{\bar k}(a,b)\}_{|\bar k|\le m}$.

 Now (\ref{tens_pr}), (\ref{res_rep}) and Definition \ref{def:1} give
\begin{multline*}
F_{2}\Big(E+\dfrac{\xi}{2n\rho(E)},E-\dfrac{\xi}{2n\rho(E)}\Big)=C_n\Big(\Big(\mathcal{K}_{*\xi,l}^{n-1}\otimes\mathcal{A}^{n-1}_m\Big)f_\xi,\bar f_\xi\Big)(1+o(1))\\
=(\mathcal{A}_m^{n-1}f_1,\bar f_1)(\mathcal{K}_{*\xi,l}^{n-1}1,1)(1+o(1)),
\end{multline*}
where we used that $f_\xi$ asymptotically can be replaced by $ f_1\otimes 1$, where $f_1$ does not depend on $\xi$ and $U_j$.
Similarly
\[
D_2=C_n(\mathcal{K}_{*0}^{n-1}\otimes\mathcal{A}_m^{n-1}f,\bar f)(1+o(1))=(\mathcal{A}_m^{n-1}f_1,\bar f_1)(\mathcal{K}_{*0,l}^{n-1}1,1)(1+o(1)),
\] 
and so
\[
\bar F_{2}\Big(E+\dfrac{\xi}{2n\rho(E)},E-\dfrac{\xi}{2n\rho(E)}\Big)=(\mathcal{K}_{*\xi,l}^{n-1}1,1)(1+o(1)),
\]
since according to Proposition \ref{p:K(U)} $\phi_0(U)=1$ is eigenvector of $K_*$ with an eigenvalue $1$, thus
\begin{equation}\label{K_*0_1}
(\mathcal{K}_{*0,l}^{n-1}1,1)= 1.
\end{equation}
Observe that the Laplace operator 
$\Delta_U$ on $U(2)$ is also reduced by $\mathcal{E}_0$ and has the same eigenfunctions as $\mathcal{K}_{*0}$ with eigenvalues 
$\lambda_j^*=j(j+1)$. 
Hence,  in the regime  $W^{-2}=C_*n^{-1}$ we can write $\mathcal{K}_{*\xi,l}$ as
\[ 
\mathcal{K}_{*\xi,l}\sim 1-n^{-1}(C^*\Delta_U+i\xi\pi \nu)\Rightarrow (\mathcal{K}_{*\xi,l}^{n-1}1,1)\to (e^{-C^*\Delta_U-i\xi\pi\hat\nu}1,1),
\]
where $C^*=C_*/t^*$, which gives Theorem \ref{thm:main}.

\section{Preliminary results}

Recall that stationary  points $X_+$, $X_-$, and $X_\pm(U)$ of the function $\mathcal{F}$ of (\ref{F_cal}) are defined in (\ref{st_points_1}).

Put
\begin{equation*}
X=
\left(
\begin{array}{cc}
a_1&(x_1+iy_1)/\sqrt{2}\\
(x_1-iy_1)/\sqrt{2}&b_1
\end{array}
\right), \quad Y=
\left(
\begin{array}{cc}
a_2&(x_2+iy_2)/\sqrt{2}\\
(x_2-iy_2)/\sqrt{2}&b_2
\end{array}
\right).
\end{equation*}
Considering the operators $K, K_\xi$ near the points $X_+$ and $X_-$, we are going to extract the contribution from the diagonal elements of $X$, $Y$.
To this end, rewrite $K(X,Y)$, $K_\xi (X,Y)$ of (\ref{K}) -- (\ref{K_xi}) as
\begin{align}
\label{rep_1}
&K_\xi (X,Y)=K(X,Y)+\widetilde K_\xi (X,Y),\\
\notag
&K(X,Y)= A(a_1,a_2)\,A(b_1,b_2)\,A_{1}(X,Y),
\end{align}
where the kernels $A$ (the contribution of the diagonal elements) is defined in (\ref{A}), and $A_1$ (the contribution of the off-diagonal elements, which however depends on diagonal
elements as well) has the form
\begin{align}
 \label{A_1}
&A_1(X,Y)=  (2\pi)^{-1}W^2F_1(X)\cdot\exp\{-W^2(x_1-x_2)^2/2-W^2(y_1-y_2)^2/2\}\cdot F_1(Y);\\ \notag
&F_1(X)=\exp\Big\{-\dfrac{1}{4}(x_1^2+y_1^2)+\dfrac{1}{2}\log\Big(1-\dfrac{x_1^2+y_1^2}{2(a_1-iE/2)(b_1-iE/2)}\Big)\Big\}. \notag
\end{align}
The perturbation kernel $\widetilde K_\xi$ in this coordinates is
\begin{align}\label{tilK_xi}
\widetilde K_\xi(X,Y)=A(a_1,a_2)\,A(b_1,b_2)\,A_{1}(X,Y)\,\Big(e^{-\frac{i}{2n\rho(E)}\big(\xi (a_1-b_1) +\xi (a_2-b_2)\big)}-1\Big).
\end{align}
It is easy to check that for $g$ defined in (\ref{A})
\begin{equation*}
g(a_{\pm}+x)-g(a_\pm)=c_{\pm}x^2+c_{3\pm}x^3+\dots
\end{equation*}
with
\begin{align}
&c_\pm=a_+(\sqrt{4-E^2}\pm iE)/2,\quad \Re c_{+}=\Re c_{-}>0,\label{c_pm}
\end{align}
and some constants $c_{3\pm}, c_{4\pm},\ldots$  

Representation of $K, K_\xi$ near  $X_\pm(U)$ was described in (\ref{rep_2}) -- (\ref{K(U)})

Following \cite{SS:ChP}, define 
 the orthonormal in $L_2[\mathbb{R}]$ system of
 functions
\begin{align}\label{pA.1}
&\psi_{0}^{\alpha}(x)=e^{-\alpha W x^2}\sqrt[4]{\alpha W/\pi},\\
&\psi_k^\alpha(x)=h_k^{-1/2}e^{-\alpha Wx^2} e^{2\Re\alpha \cdot W x^2}\Big(\frac{d}{dx}\Big)^ke^{-2\Re \alpha \cdot W x^2},
=e^{-\alpha Wx^2}p_k(x)\notag\\
&h_k^\alpha=k!(4\Re \alpha \cdot W)^{k-1/2}\sqrt{2\pi},\quad k=1,2,\ldots,
\notag\end{align}
with some $\alpha$ such that $\Re \alpha>0$,
and set
\begin{equation}\label{psi_pm}
\psi_k^\pm(x)=\psi_k^{\alpha_\pm}(x-a_\pm)
\end{equation}
with
\begin{align*}
 &\alpha_\pm=\sqrt{\dfrac{c_\pm}{2}}\Big(1+\frac{c_\pm}{2W^2}\Big)^{1/2}
\end{align*}
Now choose $W,n$-independent $\delta>0$, which is small enough to provide that the domain
$$\Omega_\delta=\{X: |\mathcal{F}(X)|>1-\delta\}$$ contains three non-intersecting subdomains $\Omega_\delta^{\pm}$, $\Omega_\delta^{+}$, $\Omega_\delta^{-}$, 
such that each of $\Omega_\delta^{+}$, $\Omega_\delta^{-}$ contains one of the points $X_+$, $X_-$, and $\Omega_\delta^{\pm}$ contains the surface $X_\pm(U)$
of (\ref{st_points_1}). 
 
 Set 
 \begin{align}\label{m}
m=[\log^2 W],
\end{align}
and consider the system of functions 
\begin{align}\label{sys_pm}
&\{\Psi_{\bar k, j,\delta}\}_{|\bar k|\le m,j\le (mW)^{1/2}},\\
&\bar k=(k_1,k_2),\,\,|\bar k|=\max\{k_1,k_2\},
\notag\end{align}
 obtained by the  Gram-Schmidt procedure from
\begin{equation*}
\{1_{\Omega_{\delta}^{\pm}}\Psi_{\bar k, j}\}_{|\bar k|\le m;j\le (mW)^{1/2}},
\end{equation*}
where
\begin{align}\label{Psi}
&\Psi_{\bar k,j}(a,b,U)=\Psi_{\bar k}(a,b)\phi_{j}(U), \\
&\Psi_{\bar k}(a,b)=\sqrt{\dfrac{2}{\pi}}\,(a-b)^{-1}\psi^+_{k_1}(a)\psi^-_{k_2}(b).
\notag
\end{align}
Similarly, consider the system of functions $\{\Psi^{+}_{\bar k,\delta}\}_{|\bar k|\le m}$ (with $\bar k=(k_1,k_2,k_3,k_4),$ $|\bar k|=\max\{k_i\}$)
 obtained by the Gram-Schmidt procedure from
\begin{equation*}
\{1_{\Omega_{\delta}^{+}}\, \psi^+_{k_1}(a)\,\psi^+_{k_2}(b) \,\psi^{+}_{k_3}(x+a_+)\,\psi^{+}_{k_4}(y+a_+)\}_{|\bar k|\le m},
\end{equation*}
 and define
$\{\Psi^{-}_{\bar k,\delta}\}_{|\bar k|\le m}$ by the same way. Denote $P_{\pm}$,  $P_+$, and $P_-$ the projections on the subspaces spanned on these three systems.
Evidently these three projection operators are orthogonal to each other. Set
\begin{align}\label{P_i}
P=P_\pm+P_++P_-,\quad\mathcal{H}_1 =P\mathcal{H}, \quad\mathcal{H}_2=(1-P)\mathcal{H},\quad \mathcal{H}=\mathcal{H}_1\oplus\mathcal{H} _2,
\end{align}
where $\mathcal{H}=L_2[\hbox{Herm}(2)]$. Besides, note that for any $\varphi$ supported in some domain $\Omega$ and any $C>0$ 
\begin{equation}\label{razm_K}
(K\varphi)(X)=O(e^{-cW^2})\,\,\hbox{for}\, X: \hbox{dist}\{X,\Omega\}\ge C>0.
\end{equation}
Now consider the operator $K$ as a block operator with respect to the decomposition (\ref{P_i}).  It has the form
\begin{align}\label{K_21.0}
&K^{(11)}=K_\pm+K_++K_-+O(e^{-cW}),\\ \notag
& K_\pm:=P_{\pm}KP_{\pm},\quad K_{+}=P_{+}KP_{+},\quad K_-:=P_{-}KP_{-},\\
&K^{(12)}=P_{\pm}K(I_{\pm}-P_{\pm})+P_{+}K(I_{+}-P_{+})+P_{-}K(I_{-}-P_{-})+O(e^{-cW}), \notag\\
& K^{(21)}=(I_{\pm}-P_{\pm})KP_{\pm}+(I_{+}-P_{+})KP_{+}+(I_{-}-P_{-})KP_{-}+O(e^{-cW}),
\notag\end{align}
where $I_{\pm}$, $I_+$, and $I_-$ are  operators of  multiplication by $1_{\Omega_{\delta}^{\pm}}$,
 $1_{\Omega_{\delta}^{+}}$, and  $1_{\Omega_{\delta}^{-}}$ respectively. Indeed,
 it is easy to see from (\ref{razm_K}) and from the relation
\[
\psi_k(x)=O(e^{-cW}) \,\,\hbox{for}\, |x|\ge C>0,\quad k\le m
\]
 that, e.g. , $P_+KP_-f=O(e^{-cW})$, $P_{\pm}K(I_+-P_+)f=O(e^{-cW})$, etc.
 
Note that  by (\ref{block})  $K_\pm$ also  has a block diagonal structure:
 \begin{align}\label{K_11.0}
&K_\pm=\sum_{j=0}^{(mW)^{1/2}}K_{\pm}^{(j)},\quad K_{\pm}^{(j)}=\mathcal{P}_jP_{\pm}KP_{\pm}\mathcal{P}_j.
\end{align}
Here and below we denote by $\mathcal{P}_ j$ the projection on $\{\Psi(a,b)\phi_{j}(U)\}$.


Let us denote by $p$ and $q$ some absolute exponents which could be different in different formulas.

Chose the contour $\mathcal{L}$ as follows:
\begin{align}
\mathcal{L}= \mathcal{L}_1\cup\mathcal{L}_2, \label{L}
\end{align}
where
\begin{align}\label{L_2}
\mathcal{L}_2=\Big\{z:|z|=|\lambda_0(K)|\Big(1-\dfrac{\log^2 W}{(a_+-a_-)^2W^2}\Big)\Big\},
\end{align}
and
\begin{align}\label{L_1}
&\mathcal{L}_1=L^0\cup L^1,\\ \notag
&L^0=\Big\{z:|z-\lambda_0(K)|= \dfrac{D^2}{(a_+-a_-)^2W^2}\Big\};\\ \notag
&L^1=\cup_{j=D}^{l-1} L_j,\quad L_j=\{z:\big |z-\lambda_{j,*}\cdot\lambda_0(K)\big|= \dfrac{\gamma}{W^2}\}
\end{align}
with
\begin{equation}\label{l}
l=\log W.
\end{equation}
Here 
\begin{align}
&\lambda_{j,*}=1-\frac{j(j+1)}{W^2(a_+-a_-)^2},
\label{lam*}
\end{align}
$\gamma>0$ and $D>0$ are sufficiently large (but $\gamma<D/2(a_+-a_-)^2$).
Notice that 
\begin{align}\label{dist1}
&\hbox{dist}\{L^{0},L^{1}\}\ge \dfrac{D}{3(a_+-a_-)^2W^2},\\ \label{dist2}
&\hbox{dist}\{\mathcal{L}_1,\mathcal{L}_{2}\}\ge \dfrac{C\log W}{W^2}.
\end{align}
Denote also
\begin{equation}\label{G_xi^0}
G_\xi^{0}(z)=(\mathcal{A}_m\otimes K_{*\xi,l}-z)^{-1},
\end{equation}
where $\mathcal{A}_m$, $K_{*\xi,l}$ are defined in (\ref{A_m}) and (\ref{K_*xi}).

We start with the following theorem
\begin{theorem}\label{t:K}
For the operators $K$ defined in (\ref{K}) we have
\begin{enumerate}

\item[(i)] For $z$ outside of the contour $\mathcal{L}$ of (\ref{L}) we have $||(K-z)^{-1}||\le CW^2$;

\item[(ii)]  Given $z$ such that $$\big|z-\lambda_{j,*}\cdot|\lambda_0(K)|\big|\ge \dfrac{\gamma}{W^2}, \quad |z|\ge |\lambda_0(K)|\Big(1-\dfrac{
\log^2W}{(a_+-a_-)^2W^2}\Big)$$ with
 sufficiently big  $\gamma>0$, consider $G^{(j)}(z)=(K_{\pm}^{(j)}-z)^{-1} $.
 Then 
\begin{align}\label{K_11.1}
& || G^{(j)}||\le C_1W^2/\gamma
\end{align}
with some absolute constant $C_1$ which does not depend on $\gamma$.

In addition, for any $z$ such that $|\lambda_0(K)|\Big(1-\dfrac{\log^2 W}{(a_+-a_-)^2W^2}\Big)\le |z|\le 1+C_2/n$
\begin{align}\label{K_11.1a}
&\|( K_+-z)^{-1}\|\le CW,\quad \| (K_--z)^{-1}\|\le CW,\quad ||(K^{(22)}-z)^{-1}||\le CW/m^{1/3}.
\end{align}

\item[(iii)] We have
\begin{align}\label{t_bound2}
&||K^{(21)}||\le Cm^{3/2}/W^{3/2},\quad ||K^{(12)}||\le Cm/W, 
\end{align}
and for $z$ outside of $\mathcal{L}$ we also have
\begin{align}
\label{t_bound3}
&||(K^{(11)}-z)^{-1}K^{(12)}||\le Cm^p,\quad ||K^{(21)} (K^{(11)}-z)^{-1}||\le Cm^p.
\end{align}
\end{enumerate}
Same statements are valid for $K_\xi$ of (\ref{K_xi}). In addition, given (\ref{G_xi^0}),
\begin{equation}\label{b_G_xi,0}
|G_\xi^0(z)|\le CW^2
\end{equation}
for $z$ outside of the contour $\mathcal{L}$.
\end{theorem}
\noindent{\it Proof of Theorem \ref{t:K}.} The proof of the theorem for $K$ and (\ref{b_G_xi,0}) follows from Lemmas 4.1 -- 4.3 and Proposition 4.1 of \cite{SS:ChP}.

To obtain the result for $K_\xi$ set
\begin{align}\label{G_1,2}
G_{1,\xi}=(K^{(11)}_\xi-z)^{-1}=(K^{(11)}+\widetilde K^{(11)}_\xi-z)^{-1},\\ \notag
G_{2,\xi}=(K^{(22)}_\xi-z)^{-1}=(K^{(22)}+\widetilde K^{(22)}_\xi-z)^{-1}.
\end{align}
Now using Schur's formula we get
\begin{align}\label{G_xi}
(K_\xi-z)^{-1}=\left(\begin{array}{cc} G_\xi^{(11)}&-G_\xi^{(11)}K^{(12)}_\xi G_{2,\xi}\\ -G_{2,\xi} K^{(21)}_\xi G_\xi^{(11)} & G_{2,\xi}+
G_{2,\xi} K^{(21)}_\xi G_\xi^{(11)} K^{(12)}_\xi G_{2,\xi}\end{array}\right),
\end{align}
where
\begin{align*}
G_\xi^{(11)} &=(K^{(11)}_\xi-z-K^{(12)}_\xi G_{2,\xi} K^{(21)}_\xi)^{-1}=(1-G_{1,\xi}K^{(12)}_\xi G_{2,\xi} K^{(21)}_\xi)^{-1} G_{1,\xi}.
\end{align*}
Denoting
\begin{equation}\label{R}
R=(1-G_{1,\xi}K^{(12)}_\xi G_{2,\xi} K^{(21)}_\xi)^{-1},
\end{equation}
we get
\begin{equation}\label{G_11}
G_\xi^{(11)} =R G_{1,\xi}.
\end{equation}
Notice that
\begin{align}\label{GK_12}
&\|G_{1,\xi}K^{(12)}_\xi\|=\|(K^{(11)}-z+\widetilde K^{(11)}_\xi)^{-1}(K^{(12)}+\widetilde K^{(12)}_\xi)\|\\ \notag
&=\|(1+(K^{(11)}-z)^{-1}\widetilde K^{(11)}_\xi)^{-1}(K^{(11)}-z)^{-1}(K^{(12)}+\widetilde K^{(12)}_\xi)\|.
\end{align}
Moreover (\ref{K_21.0}) and part (ii) of the Theorem for operator $K$ yield
\begin{equation}\label{b_K11}
\|(K^{(11)}-z)^{-1}\|\le \dfrac{C_1n}{\gamma},
\end{equation}
where $\gamma$ is sufficiently big and $C_1$ does not depend on $\gamma$. Hence
\[
\|(K^{(11)}-z)^{-1}\widetilde K^{(11)}_\xi\|\le C<1.
\]
Thus according to (\ref{GK_12}), (\ref{t_bound3}) for $K$, and (\ref{b_Ktil})
\begin{align*}
&\|G_{1,\xi}K^{(12)}_\xi\|\le C\|(K^{(11)}-z)^{-1}(K^{(12)}+\widetilde K^{(12)}_\xi)\|\\
&\le C(\|(K^{(11)}-z)^{-1}K^{(12)}\|+\|(K^{(11)}-z)^{-1}\widetilde K^{(12)}_\xi\|)\\
&\le C(\log^p W+C_1)\le C\log^p W.
\end{align*}
Similarly
\[
\|K^{(21)}_\xi G_{1,\xi}\|\le C \|(K^{(21)}+\widetilde K^{(21)}_\xi) (K^{(11)}-z)^{-1}\|\le C\log^p W.
\]
The bound (\ref{t_bound2}) for $K_\xi$ trivially follow from (\ref{t_bound2}) for operator $K$ and (\ref{b_Ktil}), which finishes the proof of  (iii) for $K_\xi$.

In addition, due to the last bound of (\ref{K_11.1a}) for operator $K$ and (\ref{b_Ktil}) we have
\begin{multline}\label{b_G22}
\|G_{2,\xi}\|=\|(K^{(22)}+\widetilde K^{(22)}_\xi-z)^{-1}\|\\
=\|(1+(K^{(22)}-z)^{-1}\widetilde K^{(22)}_\xi)^{-1}(K^{(22)}-z)^{-1}\|\le CW/m^{1/3}
\end{multline}
which gives the last bound of (\ref{K_11.1a}) for operator $K_\xi$.
This implies
\begin{equation}\label{GK_21}
\|G_{2,\xi} K^{(21)}_\xi\|\le \dfrac{\log^p W}{W^{1/2}}.
\end{equation}
Thus
\begin{equation}\label{1-R}
\|G_{1,\xi}K^{(12)}_\xi G_{2,\xi} K^{(21)}_\xi\|\le \|G_{1,\xi}K^{(12)}_\xi\|\cdot \|G_{2,\xi} K^{(21)}_\xi\|\le \dfrac{C\log^p W}{W^{1/2}},
\end{equation}
and so
\[
\|R\|\le C.
\]
This, (\ref{R}) -- (\ref{G_11}), and (\ref{b_K11}) yield
\begin{equation}\label{b_G11}
\|G_\xi^{(11)}\|\le Cn.
\end{equation}
Similarly (\ref{G_11}) gives
\begin{equation*}
\|G_\xi^{(11)}K_\xi^{(12)}\|=\|RG_{1,\xi}K_\xi^{(12)}\|\le \|R\|\cdot \|G_{1,\xi}K_\xi^{(12)}\|\le C\log^p W,
\end{equation*}
which implies
\begin{equation}\label{b_12}
\|G_\xi^{(11)}K^{(12)}_\xi G_{2,\xi}\|\le C\log^p W\cdot W.
\end{equation}
It is easy to see that 
\[
D^{-1} C (A-BD^{-1}C)^{-1}=(D-CA^{-1}B)^{-1}CA^{-1},
\]
thus
\begin{align*}
G_{2,\xi} K^{(21)}_\xi G_\xi^{(11)}&=(K^{(22)}_\xi-z-K^{(21)}_\xi G_{1,\xi}K^{(12)}_\xi)^{-1}K^{(21)}_\xi G_{1,\xi}\\
&=(1-G_{2,\xi}K^{(21)}_\xi G_{1,\xi}K^{(12)}_\xi)^{-1}G_{2,\xi}K^{(21)}_\xi G_{1,\xi}.
\end{align*}
But 
\[
\|G_{2,\xi}K^{(21)}_\xi G_{1,\xi}K^{(12)}_\xi\|\le \|G_{2,\xi}K^{(21)}_\xi \|\cdot \|G_{1,\xi}K^{(12)}_\xi\|\le \dfrac{C\log^pW}{W^{1/2}},
\]
hence using (\ref{t_bound3}) for $K_\xi$ we obtain
\begin{equation}\label{b_21}
\|G_{2,\xi} K^{(21)}_\xi G_\xi^{(11)}\|\le C\|G_{2,\xi}\|\cdot \|K^{(21)}_\xi G_{1,\xi}\|\le C\log^p W\cdot W.
\end{equation}
We also can write
\begin{equation}\label{b_22}
\|G_{2,\xi} K^{(21)}_\xi G_\xi^{(11)} K^{(12)}_\xi G_{2,\xi}\|\le \|G_{2,\xi}\|^2\cdot \|K^{(21)}_\xi \|\cdot \|G_\xi^{(11)} K^{(12)}_\xi\|\le C\log^pW\cdot W^{1/2}
\end{equation}
which finishes the proof of (i) for $K_\xi$.

Bounds (\ref{K_11.1}) -- (\ref{K_11.1a}) for $K_\xi$ can be obtained easily from those for $K$ and from (\ref{b_Ktil}).

\medskip
$\Box$
\section{Proof of Theorem \ref{thm:main}}
The key step in the proof of Theorem \ref{thm:main} is the following theorem
\begin{theorem}\label{thm:key}
Given $G_\xi(z)=(K_\xi-z)^{-1}$ with $K_\xi$ of (\ref{K_xi}), $f_\xi$ of (\ref{F_cal}), and the contour $\mathcal{L}$ defined in (\ref{L}) -- (\ref{l}), we can write for the
integral in  (\ref{res_rep}) 
\begin{multline}\label{int_rel}
\int\limits_{\mathcal{L}} z^{n-1}(G_\xi(z) f_\xi, \bar f_\xi) dz\\=\int\limits_{\mathcal{L}} z^{n-1}( G_\xi^{0}(z)(f_{1,\pm}\otimes 1), (\bar f_{1,\pm}\otimes 1)) dz +|\lambda_0(K)|^{n-1}\cdot \|f_{1}\|^2\cdot O\Big (\dfrac{1}{\log W}\Big),
\end{multline}
where 
\begin{equation}\label{f_1}
f_1=P\,f,
\end{equation}
where $P$ is the orthogonal projector to the the space $\mathcal{H}_1$ (see (\ref{P_i})), 
and $G_\xi^0$ is defined in (\ref{G_xi^0}).
Here $f_{1,\pm}$ is a projection of $f$ on the linear span of $\{\Psi_{\bar k, 0}(a,b), |k|\le m\}$ of (\ref{Psi}).

The contour $\mathcal{L}$ encircles all eigenvalues of $\mathcal{A}_m\otimes K_{*\xi,l}$ defined in (\ref{A_m}) and (\ref{K_*xi}), and
\begin{equation}\label{Af,f}
(\mathcal{A}_m^{n-1}f_{1,\pm},f_{1,\pm})=|\lambda_0(K)|^{n-1}\cdot \|f_{1}\|^2\cdot (1+o(1)).
\end{equation}
\end{theorem}
Let us assume that Theorem \ref{thm:key} is proved and derive  the assertion of Theorem \ref{thm:main}.

Indeed, since $\mathcal{L}$ encircles all eigenvalues of $\mathcal{A}_m\otimes K_{*\xi,l}$, according to the Cauchy theorem we get 
\begin{align*}
&-\dfrac{1}{2\pi i}\int\limits_{\mathcal{L}} z^{n-1}( G_\xi^{0}(z)(f_{1,\pm}\otimes 1), (\bar f_{1,\pm}\otimes 1)) dz=\big((\mathcal{A}_m\otimes K_{*\xi,l})^{n-1}(f_{1,\pm}\otimes 1), (\bar f_{1,\pm}\otimes 1)\big)\\
&=
(\mathcal{A}_m^{n-1} f_{1,\pm}, \bar f_{1,\pm})\cdot (K_{*\xi,l}^{n-1}1,1).
\end{align*}
Now 
\begin{align*}
K_{*\xi,l}=K_{*0,l}-\dfrac{i\pi\xi}{n} \nu+O(n^{-2}),
\end{align*}
where  $K_{*0,l}$ is a diagonal (in basis $\{\phi_{j}\}_{ j\le l}$ of (\ref{phi_j0})) operator with eigenvalues
$\{\lambda_{j,*}\}_{j\le l}$ of (\ref{lam*}). Since the Laplace operator 
$\Delta_U$ on $U(2)$ has the same eigenfunctions as $\mathcal{K}_{*0}$ with eigenvalues 
\[\lambda_j^*=j(j+1),\]
we get for $n=C_*W^2$
\begin{equation}\label{conv}
\mathcal{K}_{*\xi,l}\sim 1-n^{-1}(C^*\Delta_U+i\xi\pi \nu)+O(n^{-2})\Rightarrow (\mathcal{K}_{*\xi,l}^{n-1}1,1)\to (e^{-C^*\Delta_U-i\xi\pi\hat\nu}1,1),
\end{equation}
where $C^*=C_*/t^*$ as in Theorem \ref{thm:main}.

This and (\ref{Af,f}) imply that
\[
-\dfrac{1}{2\pi i}\int\limits_{\mathcal{L}} z^{n-1}( G_\xi^{0}(z)(f_{1,\pm}\otimes 1), (\bar f_{1,\pm}\otimes 1)) dz
\]
is of order
\[
|\lambda_0(K)|^{n-1}\cdot \|f_{1,\pm}\|^2,
\]
and so (\ref{int_rel}) can be rewritten as 
\begin{align*}
-\dfrac{1}{2\pi i}\int\limits_{\mathcal{L}} z^{n-1}(G_\xi(z) f_\xi, \bar f_\xi) dz=(\mathcal{A}_m^{n-1} f_{1,\pm}, \bar f_{1,\pm})\cdot (K_{*\xi,l}^{n-1}1,1) (1+o(1)),\quad n\to \infty.
\end{align*}
This, a similar relation with $\xi=0$, (\ref{F_rep}),  and (\ref{res_rep}), yield
\begin{multline*}
D_2^{-1}F_2\Big(E+\dfrac{\xi}{2n\rho(E)},E-\dfrac{\xi}{2n\rho(E)}\Big)\\=\dfrac{(\mathcal{A}_m^{n-1} f_{1,\pm}, \bar f_{1,\pm})\cdot (K_{*\xi,l}^{n-1}1,1) }{(\mathcal{A}_m^{n-1} f_{1,\pm}, \bar f_{1,\pm})\cdot (K_{*0,l}^{n-1}1,1) }(1+o(1))=(K_{*\xi,l}^{n-1}1,1)(1+o(1)).
\end{multline*}
Here we used (\ref{K_*0_1}). This relation and (\ref{conv}) complete the proof of Theorem \ref{thm:main}.
\subsection{Proof of Theorem \ref{thm:key}}

We are left to prove Theorem \ref{thm:key}. 

First we decompose $f=(f_1,f_2)$ with respect to decomposition (\ref{P_i}).
Observe that since 
\[
|\mathcal F(X)|\le 1,
\]
and $\mathcal F(X)$ exponentially decreases at $\infty$ (in eigenvalues $a,b$), we have $\|f\|=const\le 1$. Moreover it is easy to see that
\begin{align*}
\|f_1\|^2\ge \|f_{1,\pm}\|^2\ge \|\mathcal F(X)\Psi_{\bar 0,0}\|^2=\Big|(2W\Re \alpha_\pm)^{1/2}\Big(\int e^{-f(a)/2}e^{-\alpha_\pm W(a-a_+)^2}da\Big)^2\Big|^2\ge \dfrac{C}{W},
\end{align*}
with $\Psi_{\bar 0,0}$ of (\ref{Psi}). Therefore
\begin{equation}
\label{b_f1}
\|f_1\|\ge \|f_{1,\pm}\|\ge C/W^{1/2}.
\end{equation}
We start with the following simple lemma
\begin{lemma}\label{l:L_1}
The main contribution to the integral in  (\ref{res_rep}) is given by the integral over the contour $\mathcal{L}_1$ of (\ref{L_1}), i.e.
\begin{equation*}
\int\limits_{\mathcal{L}} z^{n-1}(G_\xi (z) f_\xi, \bar f_\xi) dz=\int\limits_{\mathcal{L}_1} z^{n-1}(G_\xi (z) f, \bar f) dz+|\lambda_0(K)|^{n-1}\cdot \|f_{1}\|^2\cdot 
O\Big (\dfrac{\log W}{W}\Big),
\end{equation*}
where $f$ is defined in (\ref{F_cal}). In addition,
\begin{equation}\label{int_L_2}
\int\limits_{\mathcal{L}_2} z^{n-1}(G_\xi^{0}(z)(f_{1,\pm}\otimes 1), (\bar f_{1,\pm}\otimes 1)) dz=
|\lambda_0(K)|^{n-1}\cdot \|f_{1}\|^2\cdot 
o\Big (e^{-C\log^2 W}\Big),
\end{equation}
where $\mathcal{L}_2$ is defined in (\ref{L_2}), and $G_\xi^{0}(z)$ is defined in (\ref{G_xi^0}).
\end{lemma}
\noindent{\textit{ Proof of Lemma \ref{l:L_1}.}} Since for $z\in \mathcal{L}_2$ we have
\[
|z|^{n-1}\le |\lambda_0(K)|^{n-1}\cdot e^{-C\log^2 W},
\]
we get using $\|G_\xi (z)\|\le CW^2$ (see part (i) of Theorem \ref{t:K} for $K_\xi$) that
\begin{align*}
\Big|\int\limits_{\mathcal{L}_2} z^{n-1}(G_\xi (z) f_\xi, \bar f_\xi) dz\Big|&\le C_1 |\lambda_0(K)|^{n-1}\cdot e^{-C_2\log^2 W}\cdot W^2\\
&=
|\lambda_0(K)|^{n-1}\cdot \|f_{1}\|^2\cdot 
o\Big (e^{-C\log^2 W}\Big).
\end{align*}
Here we used (\ref{b_f1}). Similarly one can obtain (\ref{int_L_2}) from (\ref{b_G_xi,0}). 

Besides, 
\begin{equation}\label{b_L}
|\mathcal{L}_1|\le C\log W/W^2, 
\end{equation}
and for $z\in \mathcal{L}_1$
\begin{equation}\label{b_zn}
|z|^{n-1}\le C|\lambda_0(K)|^{n-1}.
\end{equation}
Thus, since $\|f-f_\xi\|\le C/n$, we get according to (\ref{b_f1})
\begin{align*}
&\Big|\int\limits_{\mathcal{L}_1} z^{n-1}(G_\xi (z) (f_\xi-f), \bar f_\xi) dz\Big|\le C |\lambda_0(K)|^{n-1}\cdot  W^2\cdot \|f-f_\xi\|\cdot |\mathcal{L}_1|\\
&\le
|\lambda_0(K)|^{n-1}\cdot \dfrac{\log W}{W^2}\le |\lambda_0(K)|^{n-1}\cdot \|f_{1}\|^2\cdot 
O\Big (\dfrac{\log W}{W}\Big),
\end{align*}
which gives the lemma.

$\Box$

Lemma \ref{l:L_1} yields that we can prove (\ref{int_rel}) for $\mathcal{L}_1$ instead of $\mathcal{L}$.

The next step is to prove that we can consider only the upper-left block $K^{(11)}_\xi$ of $K_\xi$ (see (\ref{K_21.0})). More precisely, we are going to prove
\begin{lemma}\label{l:K_tr}
Given (\ref{G_1,2}) and (\ref{f_1}), we have
\begin{equation*}
\int\limits_{\mathcal{L}_1} z^{n-1}(G_\xi (z) f, \bar f) dz=\int\limits_{\mathcal{L}_1} z^{n-1}( G_{1,\xi}(z)\,f_{1}, \bar f_{1}) dz +|\lambda_0(K)|^{n-1}\cdot \|f_{1}\|^2\cdot O\Big (\dfrac{\log^p W}{W^{1/2}}\Big),
\end{equation*}
\end{lemma}
\textit{Proof of Lemma \ref{l:K_tr}.} 
According to (\ref{G_xi}) we have
\begin{align*}
&\int_{\mathcal{L}_1} z^{n-1} ((K_\xi-z)^{-1}f,\bar f) dz=\int_{\mathcal{L}_1} z^{n-1} (G^{(11)}_\xi f_{1},\bar f_{1}) dz-\int_{\mathcal{L}_1} z^{n-1} (G_\xi^{(11)}K^{(12)}_\xi G_{2,\xi} f_2,\bar f_1) dz\\
&-\int_{\mathcal{L}_1} z^{n-1} (G_{2,\xi} K^{(21)}_\xi G_\xi^{(11)} f_1,\bar f_2) dz+\int_{\mathcal{L}_1} z^{n-1} ((G_{2,\xi}+
G_{2,\xi} K^{(21)}_\xi G_\xi^{(11)} K^{(12)}_\xi G_{2,\xi}) f_2,\bar f_2) dz
\end{align*}
Thus,  we get using  (\ref{b_12}) -- (\ref{b_21}), (\ref{b_L}) -- (\ref{b_zn}), $\|f_2\|\le C$, and (\ref{b_f1})
\begin{align*}
&\Big|\int_{\mathcal{L}_1} z^{n-1} (G_\xi^{(11)}K^{(12)}_\xi G_{2,\xi} f_2,\bar f_1) dz\Big|\le \|G_\xi^{(11)}K^{(12)}_\xi G_{2,\xi}\|\cdot \|f_1\|\cdot\|f_2\|\cdot \int_{\mathcal L_1} |z|^{n-1}|dz|\\
&\le \dfrac{C\log^p W\cdot W}{W^2}\cdot |\lambda_0(K)|^{n-1}\cdot \|f_1\|\le O\Big(\dfrac{C\log^p W}{W^{1/2}}\Big)\cdot \|f_1\|^2\cdot |\lambda_0(K)|^{n-1},
\end{align*}
\begin{align*}
&\Big|\int_{\mathcal{L}_1} z^{n-1} (G_{2,\xi} K^{(21)}_\xi G_\xi^{(11)} f_1,\bar f_2)  dz\Big|\le \|G_{2,\xi} K^{(21)}_\xi G_\xi^{(11)}\|\cdot \|f_1\|\cdot\|f_2\|\cdot \int_{\mathcal L_1} |z|^{n-1}|dz|\\
&\le \dfrac{C\log^p W\cdot W}{W^{2}}\cdot |\lambda_0(K)|^{n-1}\cdot \|f_1\|\le O\Big(\dfrac{C\log^p W}{W^{1/2}}\Big)\cdot \|f_1\|^2\cdot |\lambda_0(K)|^{n-1}.
\end{align*}
Notice that $G_{2,\xi}$ of (\ref{G_1,2}) is analytic outside of $\mathcal{L}_2$ (see (\ref{K_11.1a})), and so 
\[
\int_{\mathcal{L}_1} z^{n-1} (G_{2,\xi} f_2,\bar f_2) dz=0.
\]
Hence
\begin{multline*}
\int_{\mathcal{L}_1} z^{n-1} ((G_{2,\xi}+
G_{2,\xi} K^{(21)}_\xi G_\xi^{(11)} K^{(12)}_\xi G_{2,\xi}) f_2,\bar f_2) dz\\=\int_{\mathcal{L}_1} z^{n-1} (
G_{2,\xi} K^{(21)}_\xi G_\xi^{(11)} K^{(12)}_\xi G_{2,\xi} f_2,\bar f_2) dz.
\end{multline*}
Thus (\ref{b_22}) and (\ref{b_f1}) yield
\begin{align*}
&\Big|\int_{\mathcal{L}_1} z^{n-1} (
G_{2,\xi} K^{(21)}_\xi G_\xi^{(11)} K^{(12)}_\xi G_{2,\xi} f_2,\bar f_2) dz\Big|\\
&\le \|G_{2,\xi} K^{(21)}_\xi G_\xi^{(11)} K^{(12)}_\xi G_{2,\xi}\|\cdot \|f_2\|^2\cdot \int_{\mathcal{L}_1} |z|^{n-1}|dz|\\
&\le \dfrac{C\log^p W\cdot W^{1/2}}{W^{2}}\cdot |\lambda_0(K)|^{n-1}\le O\Big(\dfrac{C\log^p W}{W^{1/2}}\Big)\cdot \|f_1\|^2\cdot |\lambda_0(K)|^{n-1}.
\end{align*}
Besides, according to (\ref{G_11}) and (\ref{1-R})
\begin{align*}
&\Big|\int_{\mathcal{L}_1} z^{n-1} ((G^{(11)}_\xi-G_{1,\xi}) f_1,\bar f_1) dz\Big|\le \|1-R\|\cdot \|G_{1,\xi}\|\cdot \|f_1\|^2\cdot \int_{\mathcal{L}_1} |z|^{n-1}|dz|\\
&\le \dfrac{C\log^pW\cdot W^2}{W^{1/2}\cdot W^2}\cdot \|f_1\|^2\cdot |\lambda_0(K)|^{n-1}=O\Big(\dfrac{C\log^p W}{W^{1/2}}\Big)\cdot \|f_1\|^2\cdot |\lambda_0(K)|^{n-1}.
\end{align*}
These bounds imply Lemma \ref{l:K_tr}.

$\Box$

Now write $K_{\xi}^{(11)}-z$, $K^{(11)}-z$ in the block form
\begin{equation}\label{M_dec}
K^{(11)}-z=\left(\begin{array}{cc}
M_{1}&M_{12}\\
M_{21}&M_{2}
\end{array}\right),\quad K_{\xi}^{(11)}-z=\left(\begin{array}{cc}
M_{1,\xi}&M_{12,\xi}\\
M_{21,\xi}&M_{2,\xi}
\end{array}\right)
\end{equation}
according to decomposition
\[
\mathcal{H}_1=\mathcal{M}_1\oplus \mathcal{M}_2,
\]
where $\mathcal{M}_1$ is a linear span of $\{\Psi_{j,k,\delta},\, \,j\le \log W, |k|\le m\}$ (see (\ref{sys_pm})). 
Then (see (\ref{K_21.0}), (\ref{K_11.0}))
\begin{align}\label{M}
&M_1=\sum_{j=0}^{\log W}K_{\pm}^{(j)},\quad K_{\pm}^{(j)}=\mathcal{P}_jP_{\pm}KP_{\pm}\mathcal{P}_j,\\ \notag
&M_2=K_++K_-+\sum_{j=\log W +1}^{(m W)^{1/2}}K_{\pm}^{(j)},\\ \notag
&M_{12}=O(e^{-cW}),\quad M_{21}=O(e^{-cW}),
\end{align}
where $\mathcal{P}_ j$ is the projection on $\{\Psi_{\bar k}(a,b)\phi_{j}(U)\}$.

Set
\begin{equation}\label{G_1,l}
G_{1,l,\xi}(z)=(K_{m,l,\xi}-z)^{-1}=(M_{1,\xi})^{-1},
\end{equation}
where $K_{m,l,\xi}$ is defined in (\ref{step2}). Notice also that, since $f_1$ does not depend on $\{U_j\}$, the part of $f_1$ corresponding to $\mathcal{M}_1$ is $f_{1,\pm}\otimes 1$.

The next step is to show 
\begin{lemma}\label{l:K_m,0}
The operator $K_\xi^{(11)}$ of (\ref{K_21.0}) can be replaced by $K_{m,l,\xi}$ of (\ref{step2}), i.e. we can write
\begin{multline*}
\int\limits_{\mathcal{L}_1} z^{n-1}(G_{1,\xi}(z) f_1, \bar f_1) dz\\
=\int\limits_{\mathcal{L}_1} z^{n-1}( G_{1,l,\xi}(z)(f_{1,\pm}\otimes 1), (\bar f_{1,\pm}\otimes 1)) dz +|\lambda_0(K)|^{n-1}\cdot \|f_{1}\|^2\cdot O\Big (\dfrac{1}{\log W}\Big).
\end{multline*}
\end{lemma}
\noindent{\it Proof of Lemma \ref{l:K_t,0}.} Denote
\begin{align*}
D_\xi=M_{1,\xi}-M_{12,\xi}M_{2,\xi}^{-1}M_{21,\xi},\quad D_{0,\xi}=1-M_{12,\xi}M_{2,\xi}^{-1}M_{21,\xi}M_{1,\xi}^{-1}
\end{align*}
and write $f_1=(f_\pm\otimes 1, f_{12})$ according to the decomposition (\ref{M_dec}).

Using Schur's formula we get
\begin{align}\label{G_1_block}
G_{1,\xi}=\left(\begin{array}{cc} D_\xi^{-1}&-D_\xi^{-1}M_{12,\xi} M_{2,\xi}^{-1}\\ -M_{2,\xi}^{-1} M_{21,\xi} D_\xi^{-1} & M_{2,\xi}^{-1}+
M_{2,\xi}^{-1} M_{21,\xi} D_\xi^{-1} M_{12,\xi} M_{2,\xi}^{-1}\end{array}\right)
\end{align}
Notice that according to (ii) of Theorem \ref{t:K} $M_{2,\xi}^{-1}$ is analytic inside of $\mathcal{L}_1$, and so
\[
\int\limits_{\mathcal{L}_1} z^{n-1}( M_{2,\xi}^{-1}  f_{12}, \bar f_{12}) dz=0,
\]
thus
\begin{align}\notag
&\int\limits_{\mathcal{L}_1} z^{n-1}(G_{1,\xi}(z) f_1, \bar f_1) dz=\int\limits_{\mathcal{L}_1} z^{n-1}( D_\xi^{-1}(f_{1,\pm}\otimes 1), (\bar f_{1,\pm}\otimes 1)) dz\\\notag-
&\int\limits_{\mathcal{L}_1} z^{n-1}( D_\xi^{-1}M_{12,\xi} M_{2,\xi}^{-1} f_{12}, (\bar f_{1,\pm}\otimes 1)) dz
-\int\limits_{\mathcal{L}_1} z^{n-1}( M_{2,\xi}^{-1} M_{21,\xi} D_\xi^{-1} (f_{1,\pm}\otimes 1), \bar f_{12}) dz\\ \label{G_1_sum}
&+\int\limits_{\mathcal{L}_1} z^{n-1}( M_{2,\xi}^{-1} M_{21,\xi} D_\xi^{-1} M_{12,\xi} M_{2,\xi}^{-1} f_{12}, \bar f_{12}) dz
\end{align}
Let $z\in \mathcal{L}_1$. Then using (\ref{K_11.0}) and (\ref{dist2}) we can write (recall that $\log W\sim \log n$)
\[
\|M_2^{-1}\|\le Cn/\log n.
\]
In addition,
\[
\|K_{\xi}^{(11)}-K^{(11)}\|\le C/n,
\]
\begin{align}\notag
&\|M_{2,\xi}^{-1}\|=\|M_2^{-1}(1+(M_{2,\xi}-M_2)M_2^{-1})^{-1}\|\le \dfrac{C_1n}{\log n}\cdot \big(1-\dfrac{C_2}{\log n}\big)^{-1}\le Cn/\log n,\\
&\|M_{12,\xi}\|\le C/n,\quad \|M_{21,\xi}\|\le C/n.\label{b_M}
\end{align}
Here we used (\ref{b_Ktil}). Part (ii) of Theorem \ref{t:K} also gives (recall $n=C_*W^2$)
\begin{align}\label{b_M1}
\|M_{1,\xi}^{-1}\|\le Cn.
\end{align}
In addition, using the resolvent identity we obtain
\begin{align}\label{M_res}
D_\xi^{-1}-M_{1,\xi}^{-1}=M_{1,\xi}^{-1}M_{12,\xi}M_{2,\xi}^{-1}M_{21,\xi}M_{1,\xi}^{-1}D_{0,\xi}^{-1}.
\end{align}
According to (\ref{b_M}) -- (\ref{b_M1}) we get
\[
\|M_{12,\xi}M_{2,\xi}^{-1}M_{21,\xi}M_{1,\xi}^{-1}\|\le C/\log n,
\]
thus
\begin{equation}\label{D_0}
\|D_{0,\xi}^{-1}\|\le C.
\end{equation}
In view of (\ref{M_res})
\[
\|D_\xi^{-1}-M_{1,\xi}^{-1}\|\le \dfrac{C n}{\log n}.
\]
Therefore, since according to (\ref{lam*}), we have for $z\in L_j$ of (\ref{L_1}) 
\[
|z|^{n-1}\le C_1|\lambda_0(K)|^{n-1}\cdot e^{-C_2 j(j+1)},
\]
and $|L_j|=2\pi\gamma/W^2 $,
we get
\begin{align*}
&\Big|\int\limits_{\mathcal{L}_1}z^{n-1}\big((D_\xi^{-1}-M_{1,\xi}^{-1})(f_{1,\pm}\otimes 1),(f_{1,\pm}\otimes 1) dz\Big|\\
&\le \dfrac{Cn}{\log n} \|f_{1,\pm}\|^2\cdot |\lambda_0(K)|^{n-1}\cdot\sum\limits_{j=D}^l |L_j|\cdot e^{-C_2 j(j+1)}\\
&\le  \dfrac{C}{\log n}\cdot \|f_1\|^2\cdot |\lambda_0(K)|^{n-1}\cdot \sum\limits_{j=D}^l  e^{-C_2 j(j+1)}\le \dfrac{C}{\log n}\cdot  \|f_1\|^2\cdot |\lambda_0(K)|^{n-1}.
\end{align*}
Now consider another integrals in (\ref{G_1_sum}). Using $D_\xi=D_{0,\xi}^{-1}M_{1,\xi}^{-1}$, we obtain similarly
\begin{align*}
&\Big|\int\limits_{\mathcal{L}_1} z^{n-1}( D_\xi^{-1}M_{12,\xi} M_{2,\xi}^{-1} f_{12}, (\bar f_{1,\pm}\otimes 1)) dz\Big|\\
&\le \dfrac{Cn}{\log n}\cdot  \|f_{1,\pm}\|\cdot \|f_{12}\|\cdot |\lambda_0(K)|^{n-1}\cdot\sum\limits_{j=D}^l |L_j|\cdot e^{-C_2 j(j+1)}\le \dfrac{C}{\log n}\cdot  \|f_1\|^2\cdot |\lambda_0(K)|^{n-1},
\end{align*}
and by the same argument
\[
\Big|\int\limits_{\mathcal{L}_1} z^{n-1}( M_{2,\xi}^{-1} M_{21,\xi} D_\xi^{-1} (f_{1,\pm}\otimes 1), \bar f_{12}) dz\Big|\le \dfrac{C}{\log n}\cdot  \|f_1\|^2\cdot |\lambda_0(K)|^{n-1},
\]
\[
\Big|\int\limits_{\mathcal{L}_1} z^{n-1}( M_{2,\xi}^{-1} M_{21,\xi} D_\xi^{-1} M_{12,\xi} M_{2,\xi}^{-1} f_{12}, \bar f_{12}) dz\Big|\le \dfrac{C}{\log^2 n}\cdot  \|f_1\|^2\cdot |\lambda_0(K)|^{n-1}.
\]
This implies the lemma.

\medskip
$\Box$

Now we have the integral
\[
\int\limits_{\mathcal{L}_1} z^{n-1}( G_{1,l}(z)(f_{1,\pm}\otimes 1), (\bar f_{1,\pm}\otimes 1)) dz.
\]
The last step is to show
\begin{lemma}\label{l:K_t,0}
The operator $K_{m,l,\xi}$ of (\ref{step2}) can be replaced by $\mathcal{A}_m\otimes K_{*\xi,l}$ (see (\ref{K_*xi}) -- (\ref{A_m})), i.e. we have
\begin{multline*}
\int\limits_{\mathcal{L}_1} z^{n-1}( G_{1,l}(z)(f_{1,\pm}\otimes 1), (\bar f_{1,\pm}\otimes 1)) dz\\
=\int\limits_{\mathcal{L}_1} z^{n-1}( G^0_\xi (z)(f_{1,\pm}\otimes 1), (\bar f_{1,\pm}\otimes 1)) dz+|\lambda_0(K)|^{n-1}\cdot \|f_{1}\|^2\cdot O\Big (\dfrac{\log^p W}{ W^{1/2}}\Big),
\end{multline*}
where $G^0_\xi$ is defined in (\ref{G_xi^0}). 
\end{lemma}
\noindent{\it Proof of Lemma \ref{l:K_t,0}.}
Using the resolvent identity we can write 
\begin{align*}
G_{1,l}(z)-G_{\xi}^0(z) =-G_{\xi}^0(z) (M_{1,\xi}-\mathcal{A}_m\otimes K_{*\xi,l})G_{1,l}(z) 
\end{align*}
Since for (\ref{pA.1})
\[
\psi_{k}^\alpha(x)= O(e^{-c\log^2W}),  \quad |x|\ge 2W^{-1/2}\log W, k\le m,
\]
we get that both $K_{m,l,\xi}$, $\mathcal{A}_m\otimes K_{*\xi,l}$ are concentrated in the $\log W/W^{1/2}$-neighbourhoods of $a_\pm$ (see \cite{SS:ChP},  for details). In this neighbourhood
\begin{align*}
&a_1-b_1=a_+-a_-+O\Big(\dfrac{\log W}{W^{1/2}}\Big),\quad a_2-b_2=a_+-a_-+O\Big(\dfrac{\log W}{W^{1/2}}\Big),\\
&t=(a_+-a_-)^2+O\Big(\dfrac{\log W}{W^{1/2}}\Big)=t_*+O\Big(\dfrac{\log W}{W^{1/2}}\Big).
\end{align*}
Thus according to (\ref{l_j})
\[
\|K_{m,l,0}-\mathcal{A}_m\otimes K_{*0,l}\|\le \dfrac{C\log W}{W^{5/2}},
\]
where  $K_{m,l,0}$, $\mathcal{A}_m\otimes K_{*0,l}$ are  $K_{m,l,\xi}$, $\mathcal{A}_m\otimes K_{*\xi,l}$ with $\xi=0$.
In addition, in this neighbourhood
\[
\|\widetilde K_\xi (X,Y) -\widetilde K_\xi(X,Y)\big|_{X=Y=X_\pm}\|\le \dfrac{C\log W}{n\sqrt{W}}.
\]
Hence, since $n\sim W^2$, we get
\[
\|K_{m,l,\xi}-\mathcal{A}_m\otimes K_{*\xi,l}\|\le \dfrac{C\log W}{W^{5/2}},
\]
and so
\begin{align*}
&\Big|\int\limits_{\mathcal{L}_1}z^{n-1}\Big((G_{1,l}(z) (f_{1,\pm}\otimes 1),(\bar f_{1,\pm}\otimes 1))-(G_\xi^0(z) (f_{1,\pm}\otimes 1),(\bar f_{1,\pm}\otimes 1)) \Big)dz\Big|\\
&\le C |\mathcal{L}_1|\cdot \dfrac{CW^4\cdot \log^p W}{W^{5/2}}\cdot \|f_1\|^2\cdot |\lambda_0(K)|^{n-1}\le \dfrac{C\log^pW}{W^{1/2}} \cdot  \|f_1\|^2\cdot |\lambda_0(K)|^{n-1}
\end{align*}
 
\medskip

$\Box$

We are left to prove (\ref{Af,f}).

According to (\ref{A_m}) and the choice of $\Psi_{\bar k}$ in (\ref{Psi}) we have
\[
\mathcal{A}_m=A_m^{(+)}\otimes A_m^{(-)}+O(e^{-c\log^2W}),
\]
where
\[
A_m^{(\pm)}=P_\pm AP_\pm,
\]
where $P_+$ and $P_-$ are the projections on the subspaces spanned on the systems
$\{\psi_{k,\delta}^+\}_{k=0}^m$ and $\{\psi_{k,\delta}^-\}_{k=0}^m$ respectively (see (\ref{psi_pm})).
The behaviour of $A_m^{(\pm)}$ was studied in \cite{SS:ChP}. In particular, it was proved in Lemma 3.3, \cite{SS:ChP} that
$|\lambda_1(A_m^{(\pm)})|\le |\lambda_0(A_m^{(\pm)})|\cdot (1-c/W)$, and so for any $g$
\[
(\mathcal{A}_m^{n-1}g,\bar g)=\lambda_0(A_m^{(+)})^{n-1}\cdot \lambda_0(A_m^{(-)})^{n-1}|(g, \Psi_{\bar 0,0})|^2 (1+o(1)).
\]
Since also $\lambda_0(K)=\lambda_0(A_m^{(+)})\cdot \lambda_0(A_m^{(-)})+O(e^{-c\log^2W})$ (see \cite{SS:ChP}, eq. (4.22)) , we get
\begin{align*}
(\mathcal{A}_m^{n-1} f_{1,\pm},\bar f_{1,\pm})=\lambda_0(K)^{n-1}\cdot |(f_{1},\Psi_{\bar 0,0})|^2(1+o(1)),
\end{align*}
where we used that $(f_{1,\pm},\Psi_{\bar 0,0})=(f_{1},\Psi_{\bar 0,0})$.

According to the definition of $\{\Psi_{\bar{k}}\}_{|\bar k|\le m}$ it is also easy to see that
\[
\|f_{1}\|^2=|(f_{1},\Psi_{\bar 0,0})|^2(1+O(1/W)).
\]
Thus 
\[
(\mathcal{A}_m^{n-1} f_{1,\pm},f_{1,\pm})=\lambda_0(K)^{n-1}\cdot\|f_{1}\|^2(1+o(1)),
\]
which completes the proof of Theorem \ref{thm:key}.

\end{document}